\documentclass[prl,twocolumn,superscriptaddress]{revtex4-2}
\usepackage[dvips]{graphicx}
\usepackage{latexsym}
\usepackage{amsmath}
\usepackage{amsfonts}
\usepackage{amssymb}
\usepackage{bm}
\begin{document}

\newcommand{\fig}[2]{\includegraphics[width=#1]{#2}}
\newcommand{{\sr}}{{Sr$_2$IrO$_4$}}
\newcommand{\etal}{\textit{et. al.}}
\newcommand{\bk}{{\textbf{k}}}
\newcommand{\bp}{{\textbf{p}}}
\newcommand{\br}{{\textbf{r}}}
\newcommand{\bx}{{\textbf{x}}}
\newcommand{\txyz}{${\bm \tau}_{xyz}$}
\newcommand{\tyz}{${\bm \tau}_{yz}$}
\newcommand{\tx}{${\bm \tau}_{x}$}

\title{Three-dimensional Stacking of Canted Antiferromagnetism and Pseudospin Current\\
 in Undoped \sr: Symmetry Analysis and Microscopic Model Realization}

\author{Yun-Peng Huang}
\thanks{These two authors contribute equally.}
\affiliation{CAS Key Laboratory of Theoretical Physics, Institute of Theoretical Physics, Chinese Academy of Sciences, Beijing 100190, China}
\affiliation{School of Physical Sciences, University of Chinese Academy of Sciences, Beijing 100049, China}

\author{Jin-Wei Dong}
\thanks{These two authors contribute equally.}
\affiliation{CAS Key Laboratory of Theoretical Physics, Institute of Theoretical Physics, Chinese Academy of Sciences, Beijing 100190, China}
\affiliation{School of Physical Sciences, University of Chinese Academy of Sciences, Beijing 100049, China}

\author{Ziqiang Wang}
\thanks{Corresponding author: wangzi@bc.edu}
\affiliation{Department of Physics, Boston College, Chestnut Hill, MA 02467, USA}

\author{Sen Zhou}
\thanks{Corresponding author: zhousen@itp.ac.cn}
\affiliation{CAS Key Laboratory of Theoretical Physics, Institute of Theoretical Physics, Chinese Academy of Sciences, Beijing 100190, China}
\affiliation{School of Physical Sciences, University of Chinese Academy of Sciences, Beijing 100049, China}
\affiliation{CAS Center for Excellence in Topological Quantum Computation, University of Chinese Academy of Sciences, Beijing 100049, China}

\begin{abstract}
Recent optical second-harmonic generation experiments observed unexpected broken spatial symmetries in the undoped spin-orbit Mott insulator \sr, leading to intensive debates on the nature of its ground state.
We propose that it is a canted antiferromagnetism with a hidden order of circulating staggered pseudospin current.
Symmetry analysis shows that a proper $c$-axis stacking of the canted antiferromagnetism and the pseudospin current lead to a magnetoelectric coexistence state that breaks the two-fold rotation, inversion, and time-reversal symmetries, consistent with experimental observations.
We construct a three-dimensional Hubbard model with spin-orbit coupling for the five localized 5$d$ Wannier orbitals centered at Ir sites, and demonstrate the microscopic realization of the desired coexistence state in a wide range of band parameters via a combination of self-consistent Hartree-Fock and variational calculations.

\end{abstract}
\maketitle

\section{I. Introduction}
The layered square-lattice iridate \sr\ has been intensively studied since the discovery of the spin-orbit Mott state \cite{Kim2008,Kim2009}, as a consequence of the interplay between spin-orbit coupling (SOC) and electron correlation\cite{Pesin2010, Witczak-Krempa2014, Rau2016, Schaffer2016, Winter2017, Hermanns2018, PhysRevB.90.041102}.
The strong SOC of Ir atoms splits the $t_{2g}$ orbitals into a fully-occupied $J_\text{eff}=3/2$ quartet and a half-filled $J_\text{eff} = 1/2$ doublet.
The latter is then localized by an otherwise moderate electronic correlation, realizing a single-band pseudospin-$1/2$ Heisenberg antiferromagnet (AFM) on the quasi-two-dimensional square lattice\cite{Kim2008}, with strong exchange couplings $J\sim 60$ meV\cite{Kim2012}.
This makes \sr\ a promising analog of the cuprates, and is thus expected to  be another platform for unconventional superconductivity \cite{PhysRevLett.106.136402, PhysRevB.89.094518, PhysRevLett.110.027002, PhysRevLett.113.177003}.
A remarkable range of cuprate phenomenology has been observed in both electron- and hole-doped \sr, including Fermi surface pockets\cite{Torre2015}, Fermi arcs\cite{Kim2014}, pseudogaps\cite{Yan2015,battisti2017universality}, and $d$-wave gaps\cite{Kim2016,Zhao2019}.
Whether a superconducting state exists as in the cuprates requires
understanding thoroughly the correlated spin-orbit entangled electronic states observed in \sr.

The ground state of the undoped \sr\ is of particular interest since it is the parent phase from which these novel spin-orbit entangled correlated states emerge.
The electron correlation in the spin-orbit Mott state results in an insulating ground state with AFM long-range order.
Neutron and resonant X-ray measurements reveal that the magnetic moments are aligned in the basal $ab$ plane, with their directions tracking the $\theta \simeq 11^\circ$ staggered IrO$_6$ octahedra rotation about the $c$ axis due to strong SOC \cite{Boseggia2013a, Crawford1994, Ye2013, Dhital2013, Torchinsky2015}.
This gives rise to a net ferromagnetic (FM) moment along the $a$ axis, in addition to the AFM component along the $b$ axis.
The net FM moment of each layer is shown to order in a $+--+$ pattern along the $c$ axis\cite{Boseggia2013, Kim2009}, where $\pm$ refers to the direction the FM moment along the $a$ axis.
A schematic illustration of the state is show in Fig. 1(a).
This magnetic ground state, hereinafter denoted as $+--+$ canted AFM (CAF), belongs to a centrosymmetric orthorhombic magnetic point group $2/m1'$ with spatial $C_{2z}$ rotation, inversion, and time-reversal symmetries\cite{Zhao2016}.
Recent optical second-harmonic generation (SHG) experiments \cite{Zhao2016} reported evidence of unexpected breaking of spatial rotation and inversion symmetries, pointing to the existence of a symmetry-breaking hidden order.
It is argued that the broken symmetries can be caused by loop-currents \cite{Zhao2016,Jeong2017,Tan2020, Murayama2021} which were proposed to account for the pseudogap physics in the high-$T_c$ cuprates \cite{Varma1997, PhysRevB.73.155113, Varma_2014}.
However, the oxygen 2$p$ states in \sr\ are much further away from ($\sim 3$ eV below) the Fermi level than those in the cuprates \cite{Kim2008, PhysRevB.74.113104}, making it disadvantageous to develop the loop-currents that requires low-energy oxygen 2$p$ states.
Furthermore, the experimental measurements \cite{Zhao2016, Jeong2017, Tan2020, Murayama2021} suggest a magnetoelectric loop-current order that is ferrocially stacked along the $c$ axis, which is incompatible with the recent observation \cite{Seyler2020} of an SHG signal that switches sign every two layers.

On the other hand, a different hidden order of circulating staggered (\textit{i.e.}, $d$-wave) pseudospin current ($d$PSCO) has been proposed \cite{Zhou2017} to describe the band dispersion and the pseudogap phenomena observed in the electron-doped \sr.
The $d$PSCO generates a $d$-wave spin-orbit density wave and gives rise to Fermi pockets and Fermi arcs in the nonmagnetic electron-doped \sr, in good agreement with angle-resolved photoemission (ARPES) and scanning tunneling microscopy (STM) measurements \cite{Torre2015, Kim2014, Kim2016, Yan2015}.
It was argued that the $d$PSCO is already present in the insulating magnetic phase of the undoped \sr, responsible for the observed splitting of the bands \cite{Torre2015} at $(\pi,0)$ whose two-fold degeneracy is otherwise protected by certain lattice symmetries \cite{Zhou2017, Han2020, Kim2021}.
While describing remarkably well the highly unconventional quasiparticle properties observed in both the electron-doped and undoped \sr, the $d$PSCO in Ref. \cite{Zhou2017} was considered in the two-dimension limit of a single IrO$_2$ layer.
Further studies on the $c$-axis stacking of the $d$PSCO and the magnetic order in realistic three-dimensional systems are necessary in order to compare directly to the findings of the nonlinear optical experiments and the interpretation in terms of intracell loop-currents.

In this work, we discuss the symmetry properties of the $c$-axis stacking of CAF, $d$PSCO, and their coexistence, and study their microscopic realization in realistic three-dimensional models for undoped \sr.
The rest of the paper is organized as follows.
In Sec. II, we perform symmetry analysis.
We find that the particular coexistence state with $+--+$ CAF and $\oplus\oplus\ominus\ominus$ $d$PSCO has the symmetries consistent with experimental observations \cite{Zhao2016, Jeong2017, Tan2020} in undoped \sr.
It is a magnetoelectric state that breaks the spatial two-fold rotation, inversion, and time-reversal symmetries.
Considering all five 5$d$ orbitals of the Ir atoms, a realistic three-dimensional tight-binding model including SOC (TB+SOC) and the structural distortion is constructed in Sec. IIIA, which describes faithfully the low-energy band structure of \sr\ with the structural distortion.
The Hubbard interactions are introduced in Sec. IIIB and treated within the Hartree-Fock approximation to account for the effects of electron correlations that generate magnetism spontaneously.
We obtain CAF phases with different $c$-axis stacking pattern self-consistently and compare their energies.
The $+--+$ CAF revealed in experiments is found to be energetically favored in a wide range of band parameters.
In Sec. IIIC, the hidden $d$PSCO is considered phenomenologically by including a variational term in the Hamiltonian.
We fix the stacking pattern of CAF to be $+--+$, and compare the energies of coexistence states with different $c$-axis stacking of $d$PSCO.
The mostly favorable stacking pattern for $d$PSCO is found to be indeed the desirable $\oplus\oplus\ominus\ominus$, supporting the above-mentioned coexistence state as the ground state of undoped \sr, with its symmetries consistent with experimental measurements.
Discussions and summaries are presented in Sec. IV.

\section{II. Symmetry analyses}

To be more precise, we denote the magnetic ground state of undoped \sr\ as $(+--+)_a$ CAF, where the subscript $a$ specifies the direction of the net FM moment, since, in principle, it can be along either $a$ or $b$ axis\cite{Porras2019,Dhital2013,Ye2013}.
Fixing the net FM moment along $\alpha=\{a, b\}$ axis, there are four possible relative stacking along the $c$ axis of the FM in-plane component of the moment in each of the four IrO$_2$ planes in a unit cell, \textit{i.e.}, $(+--+)_\alpha$, $(++--)_\alpha$, $(+-+-)_\alpha$, and $(++++)_\alpha$.
It is easy to show that there is a one-to-one correspondence between states with FM moment along $a$ axis and those with FM along $b$ axis, by performing a $C_{4z}$ rotation along the $c$ axis and a lattice translation.
Explicitly, the $(+--+)_a$, $(++--)_a$, $(+-+-)_a$, and $(++++)_a$ state are equivalent to, respectively, $(++--)_b$, $(+--+)_b$, $(+-+-)_b$, and $(++++)_b$ state.
The correspondence shall be verified numerically later in the microscopic model calculations presented in Sec. IIIB by comparing the state energies.
Therefore, without loss generality, we restrict the direction of the FM moments to be along $a$ axis and drop the subscript for the canted AFM phases in the rest of the paper, unless otherwise noted.

\begin{table}[htb]
\hspace{-2.7cm}
\textbf{(a)} Symmetries of $c$-axis stacked CAF:
\begin{ruledtabular}
\begin{tabular}{c||cccc|ccc}
stacking & $C_{2z}$ & $M_z$ & $I$ & $T$ & $C'_{2z}$ & $M'_z$ & $I'$\\
\hline
\hline
$+--+$ & \txyz & \tyz & \tx & \txyz & 0 & \tx & \tyz \\
$++--$ & \txyz & \tx & \tyz & \txyz & 0 & \tyz & \tx \\
\hline
$++++$ & $\times$ & $\times$ & \tx,\tyz & $\times$ & 0,\txyz & \tx,\tyz & $\times$\\
$+-+-$ & $\times$ & \tx,\tyz & $\times$ & $\times$ & 0,\txyz & $\times$ & \tx,\tyz \\
\end{tabular}
\end{ruledtabular}
\vspace{0.5cm}
\hspace{-2.3cm}
\textbf{(b)} Symmetries of $c$-axis stacked $d$PSCO:
\begin{ruledtabular}
\begin{tabular}{c||cccc|ccc}
stacking & $C_{2z}$ & $M_z$ &$I$ & $T$ & $C'_{2z}$ & $M'_z$ & $I'$\\
\hline
\hline
$\oplus\ominus\ominus\oplus$ & 0 & \tx & \tx & 0 & 0 & \tx & \tx \\
$\oplus\oplus\ominus\ominus$ & 0 & \tyz & \tyz & 0 & 0 & \tyz & \tyz \\
\hline
$\oplus\oplus\oplus\oplus$ & 0,\txyz & \tx,\tyz & \tx,\tyz & 0,\txyz & 0,\txyz & \tx,\tyz & \tx,\tyz \\
$\oplus\ominus\oplus\ominus$ & 0,\txyz & $\times$ & $\times$ & 0,\txyz & 0,\txyz & $\times$ & $\times$ \\
\end{tabular}
\end{ruledtabular}
\caption{Symmetries of $c$-axis stacked (a) CAF and (b) $d$PSCO. The table gives the lattice translation required for a state to recover itself after a symmetry operation of the magnetic space group $2/m1'$. Symbol $\times$ means such a lattice translation does not exist. Translation vector \tx=(1/2, 0, 0), \tyz=(0, 1/2, 1/2), and \txyz=(1/2, 1/2, 1/2) in terms of the lattice constant of the conventional unit cell shown in Fig. 1. Note that the states listed in the last two rows of both (a) and (b) are invariant under a lattice translation of \txyz, there are thus two possible lattice translations differed by \txyz. }
\vskip-0.3cm
\end{table}

The symmetries of the CAF phases (without $d$PSCO) with different $c$-axis stacking are summarized in Table I(a), which gives the lattice translation, if exist, required for a state to recover itself after a symmetry operation of the magnetic point group $2/m1'$: two-fold rotation around $z$-axis $C_{2z}$, mirror reflection about $ab$-plane $M_z$, inversion $I$, time-reversal $T$, $C'_{2z}=T C_{2z}$, $M'_z=T M_z$, and $I'=T I$.
The state does not have the corresponding symmetry if it could not recover itself by any lattice translation after a symmetry operation.
Note that, because the magnetic moments are aligned in the basal $ab$ plane, without any $c$-axis component, the time-reversal operator $T$ transforms under the same irreducible representation as $C_{2z}$, and consequently, the operators $C'_{2z}$, $M'_z$, and $I'$ are projected to identity $E$, $I$, and $M_z$, respectively, as shown in Table I(a).
The $+--+$ and $++--$ CAF states share the same symmetries and belong to the centrosymmetric orthorhombic magnetic point group $2/m1'$.
However, they are inequivalent in the presence of in-plane anisotropy \cite{Porras2019,Liu2019}, as will be shown in the microscopic model calculations presented in Sec. IIIB.
The nonmagnetoelectric $++++$ CAF breaks $\{ C_{2z}, M_z, T, I' \}$, while the magnetoelectric $+-+-$ CAF breaks $\{ C_{2z}, I, T, M'_z \}$ symmetries.
They belong to the magnetic point groups $2'/m'$ and $2'/m$, respectively.
It has been argued\cite{DiMatteo2016} that both the $++++$ and $+-+-$ CAF can potentially explain the SHG experiment \cite{Zhao2016} without invoking the loop-currents, and either of them might have been created by the laser pump used in the experiments.
The possibility of laser-induced rearrangement of the magnetic stacking, however, has been ruled out by recent comprehensive measurements \cite{Seyler2020}, which show the magnetic stacking pattern is always $+--+$ under the experimental condition before strong external field drives it to be $++++$.

\begin{table}[htb]
\begin{ruledtabular}
\begin{tabular}{c|cccc}
stacking & $C_{2z}$ & $M_z$ &$I$ & $T$ \\
\hline
\hline
$+--+/\oplus\ominus\ominus\oplus$ & $\times$ & $\times$ & \checkmark & $\times$  \\
$+--+/\oplus\oplus\ominus\ominus$ & $\times$ & \checkmark & $\times$ & $\times$  \\
$+--+/\oplus\oplus\oplus\oplus$ & \checkmark & \checkmark & \checkmark & \checkmark  \\
$+--+/\oplus\ominus\oplus\ominus$ & \checkmark & $\times$ & $\times$ & \checkmark  \\
\hline
$++--/\oplus\ominus\ominus\oplus$ & $\times$ & \checkmark & $\times$ & $\times$  \\
$++--/\oplus\oplus\ominus\ominus$ & $\times$ & $\times$ & \checkmark & $\times$   \\
$++--/\oplus\oplus\oplus\oplus$ & \checkmark & \checkmark & \checkmark & \checkmark  \\
$++--/\oplus\ominus\oplus\ominus$ & \checkmark & $\times$ & $\times$ & \checkmark  \\
\hline
$++++/\oplus\ominus\ominus\oplus$ & $\times$ & $\times$ & \checkmark & $\times$  \\
$++++/\oplus\oplus\ominus\ominus$ & $\times$ & $\times$ & \checkmark & $\times$   \\
$++++/\oplus\oplus\oplus\oplus$ & $\times$ & $\times$ & \checkmark & $\times$  \\
$++++/\oplus\ominus\oplus\ominus$ & $\times$ & $\times$ & $\times$ & $\times$  \\
\hline
$+-+-/\oplus\ominus\ominus\oplus$ & $\times$ & \checkmark & $\times$ & $\times$ \\
$+-+-/\oplus\oplus\ominus\ominus$ & $\times$ & \checkmark & $\times$ & $\times$ \\
$+-+-/\oplus\oplus\oplus\oplus$ & $\times$ & \checkmark & $\times$ & $\times$  \\
$+-+-/\oplus\ominus\oplus\ominus$ & $\times$ & $\times$ & $\times$ & $\times$  \\
\end{tabular}
\end{ruledtabular}
\caption{Symmetries of the coexistence states with possible $c$-axis stacking of CAF and $d$PSCO. }
\vskip-0.3cm
\end{table}

Before performing the symmetry analysis for the $c$-axis stacked hidden $d$PSCO, we define first the notation for its stacking pattern.
The staggered IrO$_6$ octahedra rotation about the $c$ axis results in two kinds of Ir sites, enclosed by octahedron rotated clockwise and anticlockwise, respectively, as shown in Fig. 1.
The pseudospin moments on these two sublattices are represented, respectively, by red and green arrows.
In a similar vein, the staggered rotation of IrO$_6$ octahedra gives rise to two kinds of Ir plaquettes, as depicted by the two blue squares in the $z=7/8$ and $z=3/8$ planes in Fig 1(a).
The direction of the pseudospin currents around the two kinds of Ir plaquettes is denoted by, respectively, red and green symbols at the plaquette center, with $\oplus /\ominus$ corresponding to anticlockwise/clockwise circulating pseudospin current.
The stacking of the $d$PSCO is then characterized by the red symbols in each plane, from top to bottom.
For instance, the stacking pattern for the $d$PSCO shown in Fig. 1(a) corresponds to $\oplus \oplus \ominus \ominus$.

The symmetries of the nonmagnetic phases with possible $c$-axis stacked $d$PSCO are summarized in Table I(b).
Since the $d$PSCO is invariant under the time-reversal operator $T$, any operator is identical to its product with $T$, \textit{e.g.}, $C_{2z} =C'_{2z}$, $M_z =M'_z$, and $I =I'$.
As shown in Table I(b), the $\oplus\ominus\ominus\oplus$, $\oplus\oplus\ominus\ominus$, and $\oplus\oplus\oplus\oplus$ $d$PSCO states have all the symmetries of the magnetic point group $2/m1'$, while the $\oplus\ominus\oplus\ominus$ $d$PSCO breaks mirror reflection $M_z$ and inversion $I$ but preserves the symmetries of two-fold rotation $C_{2z}$ and time-reversal $T$.
It is important to note that all of the $d$PSCO states have the two-fold rotation and time-reversal symmetries.
Thus none of them is able to describe the hidden order in \textit{hole-doped} Sr$_2$Ir$_{1-x}$Rh$_x$O$_4$ observed by SHG and polarized neutron scattering measurements.
We argue that the physics in the hole-doped \sr\ to be quite different than that on the electron-doped side.
The Rh substitution\cite{PhysRevB.86.125105, PhysRevB.89.054409, PhysRevB.92.201112, cao2016hallmarks, PhysRevB.92.081117, PhysRevB.95.060407} of the strongly spin-orbit coupled Ir in the Ir-O plane is very different than the electron doping by La substitution\cite{PhysRevLett.117.107001,PhysRevB.92.075125,Chen2018} in the off-plane charge reservoir layers or surface K doping\cite{PhysRevB.84.100402,Kim2014,Kim2016,Yan2015}.
Furthermore, a doped hole in \sr\ has a different electronic structure than that of an electron and is more likely to involve higher pseudospin states \cite{doi:10.1146/annurev-conmatphys-031218-013113}.
We therefore leave the hole-doped \sr\ aside, and consider only the undoped and electron-doped \sr.
Their unconventional low-energy quasiparticle properties observed by ARPES and STM have been described successfully by the hidden order of $d$PSCO \cite{Zhou2017}.
Our focus in this paper is to investigate the effects of $d$PSCO on the symmetry properties of the three-dimensional state, which enables a direct comparison to SHG and polarized neutron scattering experiments.
At stoichiometry where these experiments have been conducted, the N\'{e}el temperature $T_N$ and the hidden order transition temperature $T_\Omega$ are very close to each other and barely distinguishable, thus provide us unambiguously only the symmetry information of the ground state, \textit{i.e.}, the coexistence state of CAF and hidden order.

\begin{figure}
\begin{center}
\includegraphics[bb= 0 0 842 595,width=3.4in]{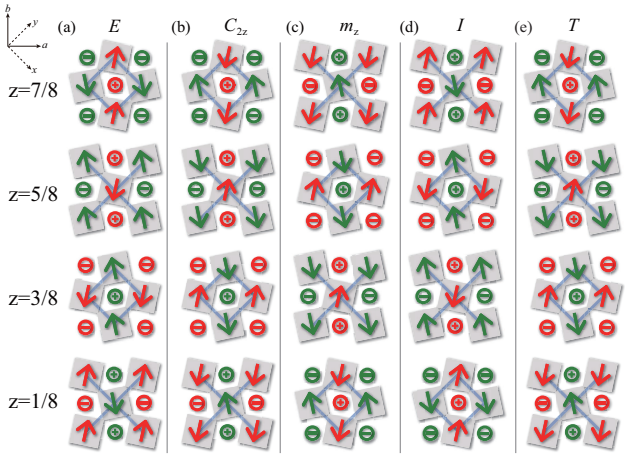}
\caption{(a) Structure of Sr$_2$IrO$_4$ with $+--+$ canted AFM and $\oplus \oplus \ominus \ominus$ $d$PSCO.
Planes through each IrO$_2$ layer in the unit cell are shown, with $z$ denoting the position of the layer along the $c$-axis.
The red and green arrows denote the direction of the magnetic moments on the two structural sublattices, and the red and blue symbols ($\oplus$ and $\ominus$) denote the direction of the pseudospin current circulating the two structural Ir plaquette.
Resultant structure upon applying the following operations contained with in the $2/m1'$ point group: (b) 180$^\circ$ rotation about the $c$-axis, (c) reflection about a mirror plane normal to the $c$-axis, (d) time-reversal, and (e) spatial inversion.
Only the structure after mirror reflection can recover the original one by a simple lattice translation.}
\label{fig1}
\end{center}
\end{figure}

The coexistence state has a symmetry only if there exist a lattice translation that simultaneously recovers both the CAF and the $d$PSCO states after the corresponding symmetry operation.
Using the symmetries of the CAF and $d$PSCO summarized in Table I, the symmetries of the coexistence states are readily obtained, with the result given in Table II for all possible $c$-axis stacking patterns.
Remarkably, there is one particular coexistence state, \textit{i.e.}, $+--+/ \oplus\oplus\ominus\ominus$ with $+--+$ CAF and $\oplus\oplus\ominus\ominus$ $d$PSCO, whose magnetism and symmetry are compatible with current experimental observations for undoped \sr.
Its magnetism is $+--+$ stacked CAF, and it breaks the two-fold rotation, inversion, and time-reversal symmetries while preserving the mirror reflection symmetry.
These properties make this coexistence state a promising candidate for the ground state of undoped \sr.
Fig. 1(a) shows the structure of the CAF and $d$PSCO in the $+--+/ \oplus\oplus\ominus\ominus$ coexistence state.
The resultant structures upon applying two-fold rotation $C_{2z}$, mirror refection $M_z$, inversion $I$, and time-reversal $T$ are shown explicitly in Fig. 1(b-e).
It is clear that only the structure in Fig. 1(c) can recover the original structure in Fig. 1(a) after a lattice translation by \tyz=(0,1/2,1/2), while the other three structures could not recover that in Fig. 1(a) by any lattice translation.

\section{III. Microscopic models}
\subsection{A. Three-dimensional TB+SOC model}
The two-dimensional TB+SOC model constructed in Ref. \cite{Zhou2017}
\begin{align}
\mathcal{H}_0 &= \sum_{ij,\mu\nu,\sigma} t_{ij}^{\mu\nu,\sigma} d^\dagger_{i\mu\sigma} d_{j\nu\sigma} +\sum_{i\mu\sigma} \epsilon_\mu d^\dagger_{i\mu\sigma} d_{i\mu\sigma} \label{H0S} \\ &+\lambda_\text{SOC} \sum_{i,\mu\nu,\sigma\sigma'} \langle \mu |\textbf{L} |\nu \rangle \cdot \langle \sigma| \textbf{S} |\sigma'\rangle d^\dagger_{i\mu\sigma} d_{i\nu\sigma'} \nonumber
\end{align}
provides a faithful description of the DFT band structure downfolded to the five low-energy Ir 5$d$-electron orbitals, as shown in Fig. 2(d).
Here, $d^\dagger_{i\mu\sigma}$ creates an electron with spin $\sigma$ at site $i$ in the $\mu$th orbital defined in the local coordinate that rotates with the IrO$_6$ octahedron, and $\mu=1 (d_{YZ})$, $2 (d_{ZX})$, $3 (d_{XY})$, $4 (d_{3Z^2-R^2})$, and $5 (d_{X^2-Y^2})$.
The crystalline electric field effects are taken into account in the on-site energy term $\epsilon_{1,\cdots,5} = (0, 0, 202, 3054, 3831)$ meV, with a separation of $\Delta_c\equiv 10Dq \simeq 3.4$ eV between the $t_{2g}$ and $e_g$ complexes.
The strength of atomic SOC $\lambda_\text{SOC}=357$ meV.
The spin and orbital angular momentum operators, \textbf{S} and \textbf{L}, have matrix elements, $S^\eta_{\sigma\sigma'}=\langle \sigma| \textbf{S} |\sigma'\rangle$ and $L^\eta_{\mu\nu} =\langle \mu |\textbf{L} |\nu \rangle$, given explicitly in Ref. \cite{Zhou2017}.
The spin-and-orbital-dependent complex hopping integrals $t^{\mu\nu, \sigma}_{ij}$ between sites $i$ and $j$ in the realistic \sr\ with structural distortion are derived from those in the idealized \sr\ without structural distortion $\tilde{t}^{\mu\nu, \sigma}_{ij}$ by transforming the $10\times 10$ hopping matrix, $t_{ij} = \mathcal{R}^\dagger_i \tilde{t}_{ij} \mathcal{R}_j$.
The operator $\mathcal{R}_i = e^{-iL_z \theta_i} \otimes e^{iS_z\theta_i}$ amounts to a joint spatial rotation from the global $(x,y,z)$ to the local $(X,Y,Z)$ coordinates by $\theta_i$ and a spin rotaion by the same angle $\theta_i$.
Note that there is a 45$^\circ$ rotation between the $x,y$ axis of the global coordinate defined in this Section and the $a,b$ axis used in Sec. II, as shown in the inset in Fig. 1.
In the undistorted idealized systems, the hopping integrals $\tilde{t}_{ij}^{\mu\nu,\sigma}$ are real, spin-independent, and given in Ref. \cite{Zhou2017} explicitly up to the fifth nearest neighbors in a IrO$_2$ layer.

\begin{figure}
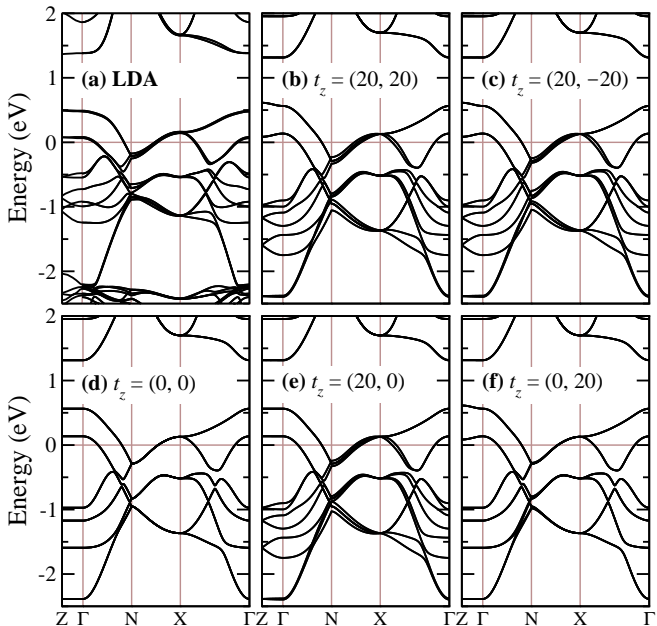

\begin{center}
\fig{4.1in}{fig2.eps}
\caption{Comparison of the band structures obtained using (a) LDA and (b-f) the three-dimensional five-orbital TB+SOC model with various inter-layer hoppings $t_z = (t_{z1}, t_{z2})$ given in meV.
The high-symmetry points labeled by $\Gamma = (0, 0, 0)$, X=$(\pi, 0, 0)$, N=$(\pi/2, \pi/2, 0)$, and Z=$(0, 0, \pi)$.}
\label{fig2}
\end{center}
\end{figure}

To construct a realistic three-dimensional model for \sr, in addition to the in-plane $\tilde{t}_{ij}$ given in Ref. \cite{Zhou2017}, we include nonzero $\tilde{t}_{ij}$'s on the nearest neighbor (NN) inter-layer bonds, \textit{i.e.}, site $i$ and $j$ from two adjacent IrO$_2$ layers.
Owing to the shift between adjacent IrO$_2$ $ab$-planes, as shown in Fig. 1(a), there are eight inter-layer NN sites for each Ir atom, four in the layer right above it and the other four in the layer right below it.
The inter-layer hoppings can be limited to the $t_{2g}$ orbitals since the contribution from the $e_g$ orbitals to the low-energy states is negligible small.
Furthermore, the inter-layer hoppings involving the planar orbital $d_{xy}$ are expected to be small.
In fact, any significant inter-layer intraorbital hopping of the $d_{xy}$ orbital would split the bands around $-2$ eV below the Fermi level, clearly incompatible with the LDA band structures shown in Fig. 2(a).
We thus consider only inter-layer hoppings involving the $d_{yz}$ and $d_{zx}$ orbitals, $t_z = (t_{z1}, t_{z2})$, where $t_{z1}$ and $t_{z2}$ denotes, respectively, the intraorbital and interorbital hoppings.
While the intraorbital hoppings are isotropic on the inter-layer NN bonds, the interorbital hoppings are anisotropic, taking values $\pm t_{z2}$ on the four inter-layer NN bonds parallel to the $[1, \pm 1, 0]$ plane in the $(x,y,z)$ global coordinates.

Figs. 2(b-f) show the electronic structures of the three-dimensional TB+SOC model with various inter-layer hoppings $t_z = (t_{z1}, t_{z2})$, in comparison to the LDA band structure plotted in Fig. 2(a).
Note that the band structures for opposite inter-layer hoppings  (\textit{i.e.}, $t_z \rightarrow - t_z$) are identical, and remain equivalent even in the presence of Hubbard interaction and $d$PSCO considered in the following subsections.
We therefore fix the intraorbital $t_{z1}$ to be positive.
Fig. 2(d) displays the band structure of the two-dimensional TB+SOC model without any inter-layer hopping, $t_z = 0$, and Fig. 2(e) and 2(f) show, respectively, the individual effects of the intraorbital $t_{z1}$ and interorbital $t_{z2}$ on the band structure.
Clearly, $t_{z2}$ does little to the band structure, while $t_{z1}$ splits the bands and thus captures the essential inter-layer features of the LDA bands displayed in Fig. 2(a).
The splitting is about $4t_{z1}$ at $N$ point for the bands right below the Fermi level.
Therefore, an intraorbital $t_{z1}$ of 20 meV would reproduce the $\sim78$ meV band splitting in the LDA bands.
The effects of the interorbital $t_{z2}$ on the band structures are negligible even in the presence of nonzero $t_{z1}$, as shown in Fig. 2(b) and 2(c).
Consequently, $t_{z2}$ remains as a tunable band parameter that shall be determined later by the $c$-axis stacking of the magnetism.

\begin{figure*}
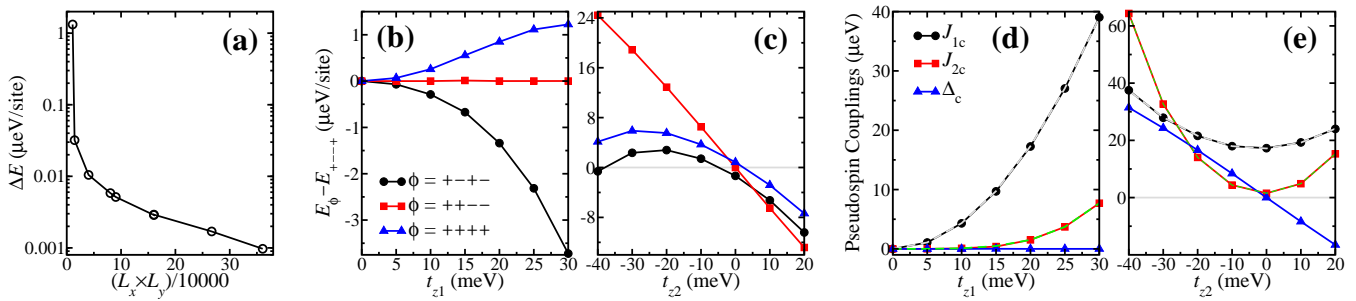

\begin{center}
\fig{7in}{fig3.eps}
\caption{(a) Energy difference between $(+--+)_a$ and $(++--)_b$ canted AFM as a function of $L_x\times L_y$, the number of in-plane \textbf{k}-points. The total number of \textbf{k}-points, $L_x\times L_y\times L_z$, in a octant of the three-dimensional reduced Brillouin zone is, for the nine data points from left to right, $121\times 90\times20$, $121\times120\times110$, $201\times200\times40$, $284\times283\times20$, $301\times300\times60$, $401\times400\times10$, $401\times400\times80$, $517\times516\times6$, and $601\times600\times10$.
The inter-layer hopping $t_z = (20, -20)$ meV.
(b) Intraorbital $t_{z1}$ dependence of the canted AFM state energies in the absence of interorbital hopping, $t_{z2}=0$.
(c) Interorbital $t_{z2}$ dependence of the canted AFM state energies with intraorbital $t_{z1}=20$ meV.
Data in (b) and (c) is obtained with $L_x\times L_y\times L_z =601\times 600\times 10$.
(d) Intraorbital $t_{z1}$ dependence of the interlayer pseudospin couplings with $t_{z2}=0$.
(e) Interorbital $t_{z2}$ dependence of the interlayer pseudospin couplings with $t_{z1}=20$ meV.
Grey/green dashed lines in (d) and (e) are fits to the data by quadratic/quartic functions.}
\label{fig3}
\end{center}
\end{figure*}

\subsection{B. Stacking of canted AFM}
To investigate the magnetism in \sr, we consider the three-dimensional five-orbital Hubbard model $\mathcal{H} = \mathcal{H}_0 + \mathcal{H}_U$, with the electron correlations described by the standard multiorbital Hubbard interactions
\begin{align}
\mathcal{H}_U = & U\sum_{i,\mu} \hat{n}_{i\mu\uparrow} \hat{n}_{i\mu\downarrow} + (U'-J_H/2) \sum_{i,\mu <\nu} \hat{n}_{i\mu} \hat{n}_{i\nu} \label{HU} \\
-&J_H \sum_{i,\mu\neq\nu} \textbf{S}_{i\mu} \cdot \textbf{S}_{i\nu} +J_H \sum_{i,\mu\neq\nu} d^\dagger_{i\mu\uparrow} d^\dagger_{i\mu\downarrow} d_{i\nu\downarrow} d_{i\nu\uparrow}, \nonumber
\end{align}
where $U$ and $U'$ are the local intraorbital and interorbital Coulomb repulsions and $J_H$ is the Hund's rule coupling with the relation of $U=U'+2J_H$.
The interactions in Eq. (\ref{HU}) are treated within the Hartree-Fock approximations.
In the presence of SOC, the Hartree and exchange self-energies induced by $\mathcal{H}_U$ depend on the full spin-orbital-dependent density matrix $n^{\mu\nu}_{i\sigma\sigma'} = \langle d^\dagger_{i\mu\sigma} d_{i\nu\sigma'} \rangle$, which are determined self-consistently in the numerical calculations.
Local physical quantities in the ground state can be expressed in terms of $n^{\mu\nu}_{i\sigma\sigma'}$, the local spin density $S^\eta_i =\sum_{\mu,\sigma\sigma'} S^\eta_{\sigma\sigma'} n^{\mu\mu}_{i\sigma\sigma'}$, and the local orbital angular momentum $L^\eta_i =\sum_{\mu\neq\nu, \sigma} n^{\mu\nu}_{i\sigma\sigma} L^\eta_{\mu\nu}$.
In all calculations presented in this paper, we choose $(U, J_H)=(1.2, 0.05)$ eV that, in the two-dimensional calculations \cite{Zhou2017},  produces correctly the CAF as the magnetic ground state for the undoped \sr, and the low-energy quasiparticle properties in good agreement with ARPES measurements \cite{Torre2015}.

We first verify numerically the one-to-one correspondence, discussed in the previous section, between the CAF states with FM moment along $a$ axis and those with FM moment along $b$ axis.
The direction of the net FM moment can be pinned by choosing appropriate initial values for $n^{\mu\nu}_{i\sigma\sigma'}$.
In numerical calculations, an octant of the reduced Brillouin zone, corresponding to the conventional unit cell with eight Ir atoms shown in Fig. 1(a), is discretized evenly into $L_x\times L_y\times L_z$ \textbf{k}-points.
We obtain these states self-consistently at various $L_x\times L_y\times L_z$ and compare their energies.
Fig. 3(a) plots the energy difference between the $(+--+)_a$ and $(++--)_b$ CAF states as a function of the in-plane \textbf{k}-point $L_x\times L_y$, with the inter-layer hopping fixed to be $t_z =(20, -20)$ meV.
The energy difference is not sensitive to $L_z$, probably because the inter-layer hoppings $t_z$ are much smaller in amplitude than the in-plane hoppings.
Except for the first two data points, the energy difference is less than 0.01 $\mu$eV per site, within the resolution of our numerical calculations.
We thus conclude that the $(+--+)_a$ and $(++--)_b$ canted AFM states are equivalent, consistent with the symmetry analysis.
The correspondences between other states are also verified numerically.
To reduce the finite-size effect, we take $L_x\times L_y\times Lz =601\times 600\times 10$ in the rest of the paper.

\begin{figure*}
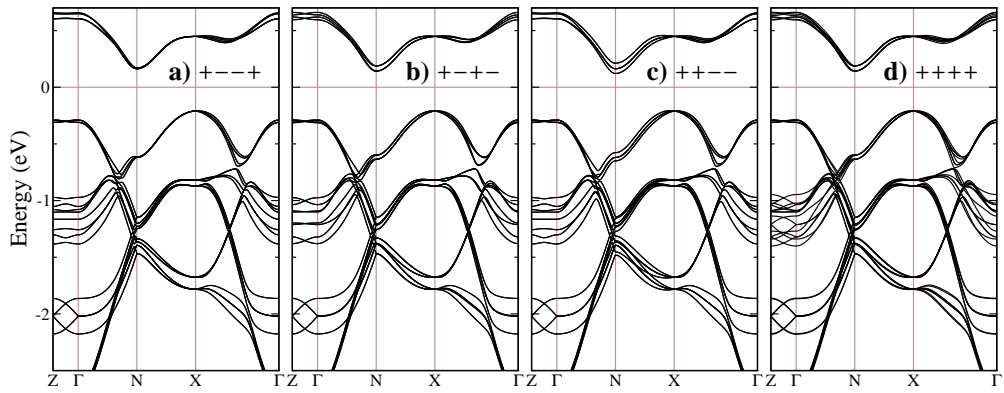

\begin{center}
\fig{5.2in}{fig4.eps}
\caption{Comparison between the band dispersions of the CAF states with various $c$-axis stacking. The inter-layer hoppings $t_z = (20, -20)$ meV and the Hubbard interactions $(U, J)=(1.2, 0.05)$ eV.}
\label{fig4}
\end{center}
\end{figure*}

In the absence of interorbital hopping, $t_{z2}=0$, the intraorbital $t_{z1}$ dependence of the state energy of $+-+-$, $++--$, and $++++$ CAF is shown in Fig. 3(b), with respect to that of the $+--+$ CAF.
Clearly, the $+--+$ and $++--$ CAF are identical in energy at any intraorbital $t_{z1}$, implying the absence of the in-plane anisotropy.
It is thus equivalent for the net FM moment to align in either the $a$-axis or in the $b$-axis, in the absence of $t_{z2}$.
The intraorbital $t_{z1}$ energetically favors the $+-+-$ CAF, while disfavors mostly the $++++$ CAF.
Fixing intraorbital $t_{z1}=20$ meV by the $\sim 78$ meV band splitting in LDA, Fig. 3(c) plots the energies of the CAF states as a function of interorbital $t_{z2}$.
While the $+--+$ CAF is the most unfavored magnetic state on the positive $t_{z2}$ side, there is a wide range on the negative side, $t_{z2} \in (-38, -6)$ meV, where the $+--+$ CAF becomes the lowest in energy, supporting the ground magnetic structure revealed in experiments \cite{Boseggia2013, Kim2009}.
Furthermore, the $++++$ CAF is higher in energy by about 5 $\mu$eV per site at $t_{z2}=-20$ meV, which agrees remarkably well with the $\sim$0.3 T external magnetic field required in experiments to align all FM moment along one direction\cite{Kim2009}.

Within the pseudospin-only models \cite{Porras2019, PhysRevB.94.224420}, it has been shown that the $c$-axis stacking of the static long-range magnetic order in the undoped \sr\ is stabilized by the interplay of interlayer pseudospin couplings, including the first-nearest-interlayer interaction $J_{1c}$, the second-nearest-interlayer interaction $J_{2c}$, and the anisotropy $\Delta_c$ comes from the anisotropic interlayer interaction\cite{Katukuri2014}.
In terms of the effective couplings between the net moments ($\tilde{\bf S}={\bf L} +2{\bf S}$), $j_{1c}=4J_{1c}\tilde{S}^2 \sin^2 \theta$, $j_{2c}=-J_{2c}\tilde{S}^2 (\cos^2\theta -\sin^2\theta)$, and $\delta_c =4\Delta_c \tilde{S}^2 \cos^2\theta$, the energies of the $+--+$, $++--$, $+-+-$, and $++++$ CAF states are, respectively, $-\delta_c -j_{2c}$, $\delta_c -j_{2c}$, $-j_{1c}+j_{2c}$, and $j_{1c}+j_{2c}$.
In the CAF states obtained self-consistently at Hubbard interactions $(U, J)=(1.2, 0.05)$ eV, the ordered pseudospin moment $\tilde{S} \simeq 0.67$ $\mu_B$ and the canting angle $\theta \simeq 22^\circ$.
It is readily and instructive to extract, from the Hartree-Fock state energies given in Fig. 3(b) and 3(c), the values of $J_{1c}$, $J_{2c}$, and $\Delta_c$.
The interlayer pseudospin couplings are plotted in Fig. 3(d) as a function of intraorbital $t_{z1}$ in the absence of interorbital $t_{z2}=0$, and in Fig. 3(e) as a function of interorbital $t_{z2}$ with the intraorbital hopping fixed to be $t_{z1}=20$ meV.
Clearly, the intraorbital $t_{z1}$ does not generate any anisotropy $\Delta_c$, while the interorbital $t_{z2}$ produces an anisotropy linear in $t_{z2}$.
In the absence of $t_{z2}$, the intraorbital $t_{z1}$ produces a $J_{1c} \propto t^2_{z1}$ and a $J_{2c} \propto t^4_{z1}$, as shown in Fig. 3(d).
At a fixed nonzero intraorbital $t_{z1}$, the superexchange interactions $J_{1c}$ and $J_{2c}$ generated by interorbital $t_{z2}$ can be well fitted by quadratic and quartic functions of $t_{z2}$ respectively, as shown in Fig. 3(e).
These behaviors are consistent with the fact that superexchange interactions $J_{1c}$ and $J_{2c}$ are generate by, respectively, the second-order and quadratic-order perturbations in the inter-layer hoppings.
Interestingly, at $t_z=(20, -20)$ meV, the interlayer pseudospin couplings $(J_{1c}, J_{2c}, \Delta_c )=(21.6, 14.0, 16.6)$ $\mu$eV are consistent with the values extracted from experiment \cite{Porras2019}.

Fig. 4 displays the band dispersions of the CAF states with different $c$-axis stackings.
They are all AFM insulators with a similar overall structure.
The quasiparticle band below the AFM gap has an eight-fold degeneracy at $X$ point, four of them due to the folding along the $c$ axis of the conventional unit cell and the other two are protected by the two-fold rotation symmetry $C_{2a}$ about the $a$ axis, along which the FM moment aligned.
These eight bands behave differently along the $X$-$N$ direction, which is probably the most pronounced difference among these band structures.
They remain degenerate in the $+--+$ CAF, split into three branches in the $++--$ CAF, but split into two branches in the $+-+-$ and $++++$ CAF states.

\subsection{C. Stacking of $d$PSCO}
The physical origin of the $d$PSCO is still under investigation \cite{Zhou2017}, and out of the scope of the current paper.
Therefore, unlike the CAF, its stacking pattern could not be determined self-consistently by including in the Hamiltonian an interaction term from which the $d$PSCO develops spontaneously.
Instead, we determine its $c$-axis stacking via a variational approach.
Explicitly, a variational term for the $d$PSCO, $\mathcal{H}_\Delta$, is added to the Hamiltonian, $\mathcal{H}=\mathcal{H}_0 +\mathcal{H}_U +\mathcal{H}_\Delta$, with
\begin{equation}
\mathcal{H}_\Delta = i\Delta \sum_{i\in A,\sigma} \sum_{j=i+\delta} \eta_i \tau_{ij}  \sigma \left( \gamma^\dagger_{i\sigma} \gamma_{j\sigma} -\chi^\sigma_{ij} \right) +\text{H.c.}, \label{HD}
\end{equation}
where the NN vector $\delta= \{\pm \hat{x}, \pm \hat{y}\}$, the standard NN $d$-wave form factor $\tau_{ij} =(-1)^{i_y+j_y}$, and $\chi^\sigma_{ij} = \langle \gamma^\dagger_{i\sigma} \gamma_{j\sigma} \rangle$ whose presence ensures that the variational term $\mathcal{H}_\Delta$ does not add an elastic part to the state energy.
The operator $\gamma_\sigma = \frac{1}{\sqrt{3}} ( i\sigma d_{yz,\bar\sigma} + d_{zx,\bar\sigma} +id_{xy,\sigma} )$
annihilates the $J_\text{eff} = 1/2$ doublet in the quasiparticle excitations, $|J=1/2, J_z = \pm 1/2 \rangle = \gamma^\dagger_\pm |0\rangle$.
The $c$-axis stacking of $d$PSCO is then controlled by $\eta_i$ as it takes on values of $\pm 1$ for Ir site $i$ in different IrO$_2$ layers.
For example, to generate the $\oplus\oplus\ominus\ominus$ stacking pattern for $d$PSCO, $\eta_i$ take the value of $+1$, $+1$, $-1$, and $-1$, respectively, for lattice site $i$ in the four IrO$_2$ layers.
We fix the the stacking pattern of CAF to be $+--+$, and try to find the energetically preferred stacking pattern of $d$PSCO in the coexistence state.

\begin{figure*}
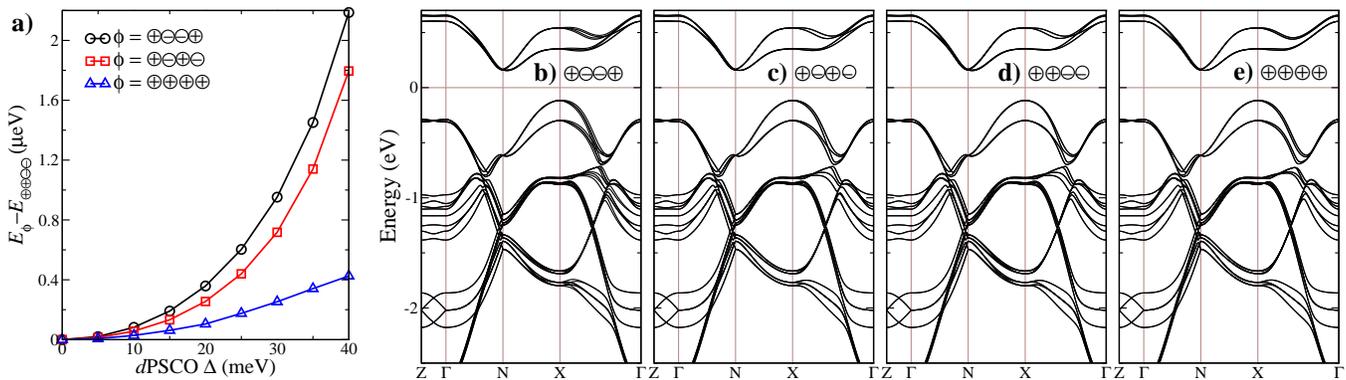

\begin{center}
\fig{7in}{fig5b.eps}
\caption{(a) The state energies of the coexistence states as a function of the $d$PSCO order $\Delta$. The stacking pattern of the canted AFM is fixed to be $+--+$, and the interlayer hopping $t_z = (20, -20)$ meV. (b-e) Comparison of the band dispersions of the coexistence states with various $c$-axis stacking of $d$PSCO ($\Delta = 30$ meV).}
\label{fig5}
\end{center}
\end{figure*}

The interlayer hoppings are chosen to be $t_z = (20, -20)$ meV where the $+--+$ CAF is the magnetic ground stare.
At a given strength of $d$PSCO, $\Delta$, we obtain the coexistence states of $+--+$ CAF with four possible $c$-axis stacked $d$PSCO self-consistently, and then compare their energies to find the preferred stacking pattern for the $d$PSCO.
The energies of the coexistence states with $\oplus \ominus \ominus \oplus$, $\oplus \ominus \oplus \ominus$, and $\oplus \oplus \oplus \oplus$ $d$PSCO are plotted in Fig. 5(a) as a function of $\Delta$, with respect to that of the coexistence state with $\oplus \oplus \ominus \ominus$ $d$PSCO.
It is clear that the $+--+/ \oplus \oplus \ominus \ominus$ coexistence state is energetically favored over all other coexistence states and, as shown in Sec. II, its symmetry is compatible with available experimental observations on undoped \sr\ below the N\'{e}el temperature $T_N$.

The band dispersions of the four coexistence states with $+--+$ CAF  are plotted in Figs. 5(b-e) for $\Delta =30$ meV.
The $d$PSCO order breaks the $C_{2a}$ symmetry, splits the eight-fold degenerate band at $X$ point into two four-fold degenerate branches, giving rise to a band splitting $\sim200$ meV at $X$ point.
We note that the $C_{2a}$ symmetry is broken by the staggered tetragonal distortion of the IrO$_6$ octahedra at temperatures above $T_\Omega$ in undoped \sr\ \cite{Ye2013, Dhital2013, Boseggia2013, Torchinsky2015}.
However, without the important $d$-wave form factor, the tetragonal distortion is unable to capture the unconventional quasiparticle properties of \sr\ in both the undoped magnetic insulating phase and the electron-doped nonmagnetic phase.

\section{IV. Discussions and summaries}
The existence and the nature of a hidden order in \sr\ have been under intensive debate.
After the observation of the anomalous SHG signal \cite{Zhao2016}, alternative explanations without invoking loop currents were subsequently proposed by Matteo and Norman \cite{DiMatteo2016}, including laser-induced rearrangement of the magnetic stacking and enhanced sensitivity to surface rather than bulk magnetism.
Polarized neutron diffraction \cite{Jeong2017} and muon spin relaxation \cite{Tan2020} measurements performed on undoped and hole-doped \sr\ revealed broken time-reversal symmetry below  $T_\Omega$.
Meanwhile, a resonant X-ray scattering measurement \cite{Chen2018} conducted on the electron-doped \sr\ has uncovered an incommensurate magnetic scattering in the pseudogap phase.
These experimental observations support the idea that the pseudogap is associated with a symmetry-breaking hidden order.
Recently, comprehensive experiments \cite{Seyler2020} conducted on undoped \sr\ have ruled out the possibility of laser-induced rearrangement of the magnetic stacking, and suggest that the surface-magnetization induced electric-dipole process in the SHG experiments can be strongly enhanced by SOC.
However, the existence of a hidden order in \sr\ remains as a possible explanation for the experimental observations of symmetry breaking and unconventional quasiparticle excitations.

In this work, we have shown that the coexistence state of $+--+$ CAF and $\oplus \oplus \ominus \ominus$ $d$PSCO has all the symmetry properties compatible with the available experimental observations on the undoped \sr\ below the N\'{e}el temperature $T_N$.
It is a magnetoelectric state that breaks the two-fold rotation $C_{2z}$, inversion $I$, and time-reversal $T$ symmetries.
We then demonstrated its microscopic realization in a three-dimensional Hubbard model with spin-orbit coupling.
Together with the fact that the highly unconventional quasiparticle properties observed in both the parent and electron-doped \sr\ can be described remarkably well by the $d$PSCO \cite{Zhou2017}, the latter offers a promising candidate for the hidden order responsible for the pseudogap phase in the undoped and electron-doped iridates.
The N\'{e}el temperature $T_N$ and the hidden order transition temperature $T_\Omega$ are very close to each other in undoped \sr. As a result, available experiments on undoped \sr\ could not tell us unambiguously the symmetry properties of the pseudogap phase.
It is thus very desirable to carry out the optical SHG and neutron scattering experiments on the pseudogap phase in electron-doped \sr.
In the absence of magnetism, the $d$PSCO would preserve the spatial inversion and time-reversal symmetries, but lower the four-fold rotation symmetry to two-fold $C_{2z}$ due to the $d$-wave form factor.

\section{V. Acknowledgments}
YH, JD, and SZ are supported by the Strategic Priority Research Program of CAS (Grant No. XDB28000000) and the National Natural Science Foundation of China (Grants No. 11974362 and No. 12047503). ZW is supported by the U.S. Department of Energy, Basic Energy Sciences (Grant No. DE-FG02-99ER45747). Numerical calculations in this work were performed on the HPC Cluster of ITP-CAS.

\end{document}